\documentclass[aps,pre,twocolumn,galleyproof,superscriptaddress]{revtex4-1}
\usepackage{amsmath,amssymb}
\usepackage{graphicx,color,dsfont}
\usepackage{etoolbox}
\usepackage[titletoc,toc]{appendix}

\usepackage{lineno} 

\DeclareGraphicsExtensions{.pdf,.png,.jpg}
\begin{document}

\title{Daily rhythms in mobile telephone communication}
\date{\today}

\author{Talayeh Aledavood}\affiliation{Aalto University School of Science, P.O. Box 12200, FI-00076, Finland}
\author{Eduardo L\'opez}\affiliation{CABDyN Complexity Center, Sa\"id Business School, University of Oxford, Oxford OX1 1HP, United Kingdom}
\author{Sam G. B. Roberts}\affiliation{Department of Psychology, University of Chester, Chester CH1 4BJ, United Kingdom}
\author{Felix Reed-Tsochas}\affiliation{CABDyN Complexity Center, Sa\"id Business School, University of Oxford, Oxford OX1 1HP, United Kingdom}\affiliation{Department of Sociology, University of Oxford, Oxford OX1 3UQ, United Kingdom}
\author{Esteban Moro}\affiliation{Departamento de Matem\'{a}ticas \& GISC, Universidad Carlos III de Madrid, 28911 Legan\'{e}s Spain}
\author{Robin I. M. Dunbar}\affiliation{Department of Experimental Psychology, University of Oxford, Oxford OX1 3UD, United Kingdom} 
\author{Jari Saram\"aki}\affiliation{Aalto University School of Science, P.O. Box 12200, FI-00076, Finland}

\begin{abstract}

Circadian rhythms are known to be important drivers of human activity and the recent availability of electronic records of human behaviour has provided fine-grained data of temporal patterns of activity on a large scale. Further, questionnaire studies have identified important individual differences in circadian rhythms, with people broadly categorised into morning-like or evening-like individuals. However, little is known about the social aspects of these circadian rhythms, or how they vary across individuals. In this study we use a unique 18-month dataset that combines mobile phone calls and questionnaire data to examine individual differences in the daily rhythms of mobile phone activity. We demonstrate clear individual differences in daily patterns of phone calls, and show that these individual differences are persistent despite a high degree of turnover in the individuals' social networks. Further, women's calls were longer than men's calls, especially during the evening and at night, and these calls were typically focused on a small number of emotionally intense relationships. These results demonstrate that individual differences in circadian rhythms are not just related to broad patterns of morningness and eveningness, but have a strong social component, in directing phone calls to specific individuals at specific times of day.   
\end{abstract}
\maketitle


\section{Introduction}

Human activity follows a circadian rhythm that is reflected at the psychological, physiological and biochemical 
levels~\cite{Kerkhof1985, Czeisler1999, Panda2002}. This rhythm is driven by endogenous cellular mechanisms, but may be modulated by 
exogenous factors. Circadian rhythms are in general synchronized to the day-night cycle. Within this cycle, there are differences between individuals. Notably, there are morning and evening types, those who wake up early and those who prefer to sleep late. This may result from intrinsic differences in the circadian pacemaker circuit; morningness and eveningness have also associated with gender and personality traits~\cite{Zhao2014, Tsaousis2010, Keren2010}. Circadian patterns can be also detected in statistics aggregated over large numbers of individuals. They have been observed in a wide range of phenomena: in human suicidal acts, times of exhibiting unethical behaviour, times of sexual activity, and times of heart attacks~\cite{Preti2001, Kouchaki2014, Refinetti2005, Hu2004}.
 
Aggregate-level daily rhythms are known to appear in electronic records of human activity, from mobility~\cite{Song2010}
 to Wikipedia and OpenStreetMap edits~\cite{Yasseri2012, Yasseri2013}, activity on Twitter~\cite{Thij2014}, and the number of mobile phone calls per hour~\cite{Ho2012, Louail2014}. In addition to the day-night cycle, these patterns 
 are modulated by a number of endogenous factors such as the daily work schedule, commuting patterns, and the activity patterns 
 of one's social circles. In studies of daily rhythms inferred from electronic records, the focus has typically remained at the aggregate level (e.g. the rhythms of cities in~\cite{Louail2014}). 
 
 Thus previous work in this area has either used questionnaires to examine broad patterns of morningness and eveningness~\cite{Zhao2014, Tsaousis2010, Keren2010}, or used electronic records to examine circadian rhythms in more detail, aggregated over many individuals. In this paper, we use a unique longitudinal dataset~\cite{Roberts2011,Saramaki2014} that combines questionnaire data with mobile phone records, allowing us to take an individual-centric point of view, and study in detail the daily rhythms of mobile telephone calls of individuals.  Our aim is to go beyond observing circadian rhythms in call frequency alone. To this end, we look at three questions: First, are there clear individual differences in daily call frequency patterns, and if so, how persistent are these given network turnover i.e. as network members change, does the daily call pattern also change? Second, are there daily patterns in other observables than call rates -- specifically, the social aspects of call diversity and who is being called? Are calls made more randomly at certain times of day and directed at certain contacts at other times? Third, do the properties of callers and callees, such as gender, explain some features of the observed patterns?
 
To this end, we employ a data set that comprises the exact times and recipients of all outgoing mobile phone calls of 24 individuals ("egos" in the following) for 18 months. This data set consists of 74,124 phone calls altogether. This set was originally collected for the purpose of studying social network turnover over time~\cite{Roberts2011}: during the study, the participants finish high school and go to work or university (often in another city), which gives rise to major turnover in their personal networks. The mobile phone call data set is accompanied by 3 surveys on the contacts who were called ("alters" in the following), including their gender and information on kinship.

Together, these data allow us to study the daily rhythms of calls in terms of numbers and in terms of recipients (who is called and when). We find that in terms of call frequency at each hour of day, each individual has their distinct, persistent pattern. These daily patterns persist for individuals despite a high degree of network turnover, and thus appear to be characteristics of individual egos, rather than dependent on the identify of specific alters. Within these patterns, there are clear variations in the entropy of called alters, indicating that certain times of day (evening and night, typically) are reserved for calling specific alters, whereas at other times the recipients of calls are more diverse. For female egos, these variations are accompanied by strong variations in the average duration of calls that increases towards the night -- these long calls are typically made to friends instead of family members.
%
%
%

\begin{figure*}[!ht]
\includegraphics[width=0.8\textwidth]{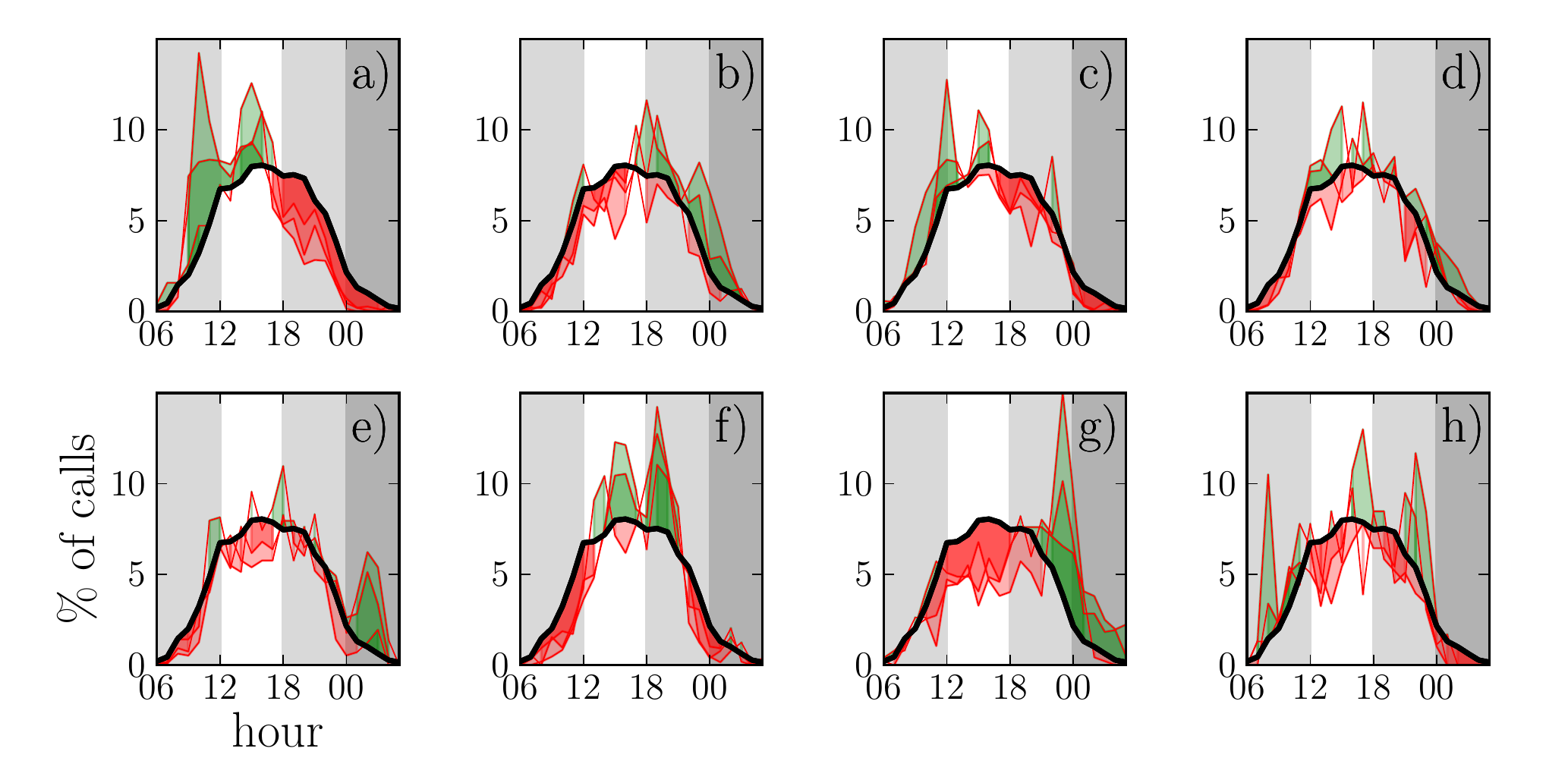}
\caption{The daily call patterns of 8 individuals (a-h). The red lines denote the average fraction of calls placed at the corresponding hour for each of  the three intervals $I_1$, $I_2$, and $I_3$. The black line is the average call pattern of all 24 individuals over all intervals. Areas shaded green show
where an individual's fraction of calls exceeds the average, while areas shaded red show where it falls below the average.}
\label{fig:daily_patterns}
\end{figure*}
\section{Results}

\subsection{Persistence of individual daily call patterns}

We begin by computing the daily call patterns for all 24 egos. The data time span is divided to three consecutive 6-month intervals $I_1$, $I_2$, and $I_3$.  For each ego and each 6-month interval, we compute the average fraction of calls placed at each hour of the day. Considering 6-month intervals separately allows investigation the persistence of any observed differences: were specific features of individual patterns due to random fluctuations alone, they would not persist over all intervals.

The resulting daily call patterns for 8 representative egos (4 male, 4 female) for all intervals are displayed in Fig.~\ref{fig:daily_patterns}. Two features clearly stand out: First, while the call patterns of all egos follow the day-night cycle and calls at night are infrequent, there are significant differences between individuals. As an example, the ego whose pattern is displayed in panel a) makes more calls in the morning than others, whereas for the ego of panel g) there are frequent calls at late hours. Second, it appears that each individual's specific patterns are rather similar in all 6-month intervals. Both observations hold for all 24 egos. This persistence is noteworthy, since it is known that at the same time, the social networks of these individuals undergo major turnover~\cite{Saramaki2014}. Because of this, the observed persistence points towards intrinsic driving forces behind the daily patterns, as these do not strongly depend on an ego's personal network composition. 

\begin{figure}[b]
\includegraphics[width=0.95\linewidth]{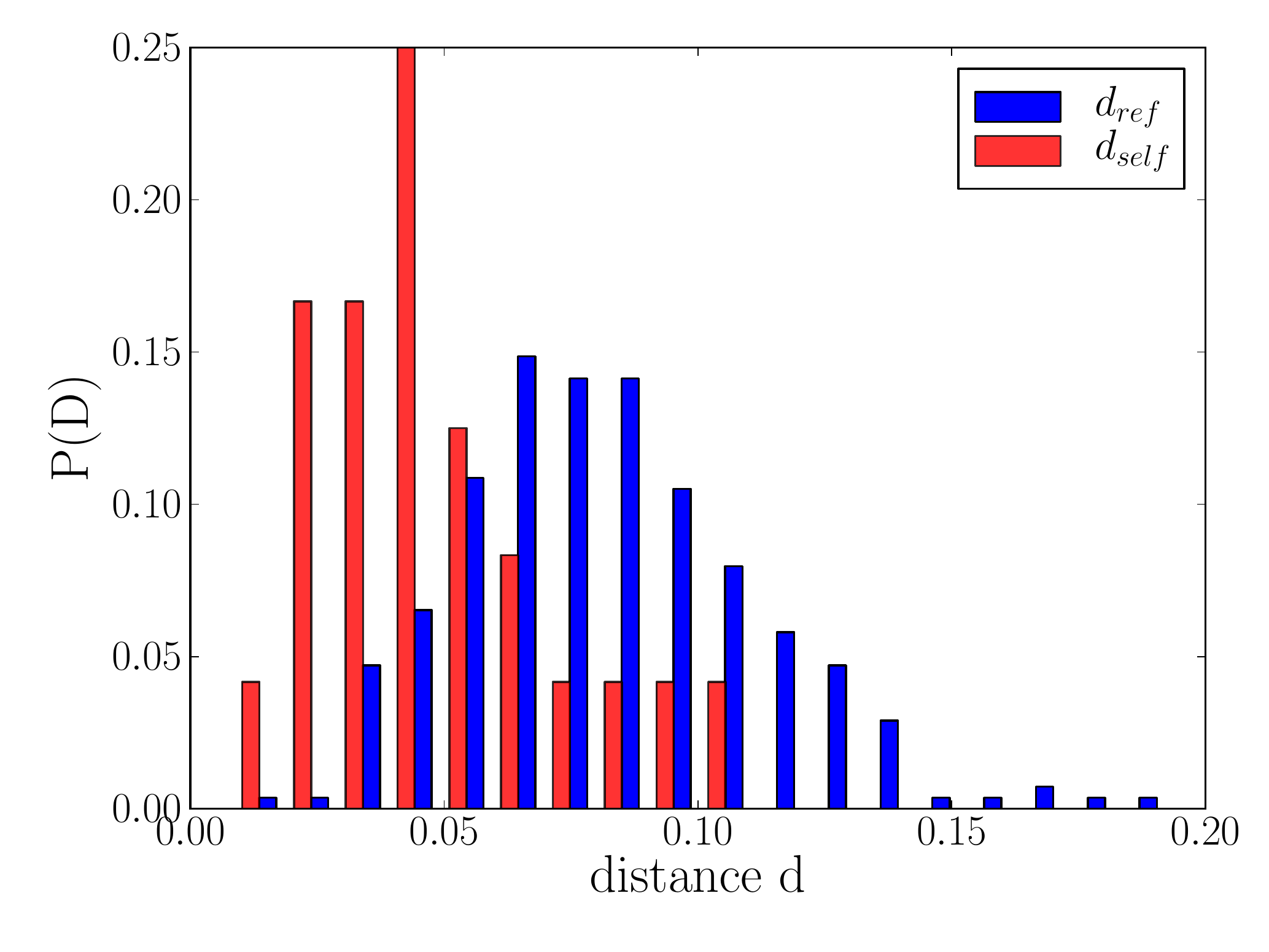}
\caption{Histogram of $d_\mathrm{self}$ and $d_\mathrm{ref}$ calculated for each ego. This plot shows the results for all egos and all time intervals. }
\label{fig:JSD}
\end{figure}

The persistence of individual daily patterns is confirmed with a more detailed analysis. Here we use the approach of Ref.~\cite{Saramaki2014} to show that  the daily call patterns of an individual in different time intervals are more similar than the patterns of different egos within one time interval. We use the Jensen-Shannon divergence (JSD) (see Appendix for details) to measure the difference between daily call patterns. For each ego, we calculate two different distances: self ($d_\mathrm{self}$) and reference ($d_\mathrm{ref}$). The self-distance $d_\mathrm{self}$ for an individual $i$ is the average JSD between the call patterns in ($I_1,I_2$) and ($I_2,I_3$): $d_{i,\mathrm{self}} = \frac{1}{2}\left(d_{12}^{i,\mathrm{self}}+d_{23}^{i,\mathrm{self}}\right)$. The reference distance measures the divergence of patterns of different egos in one time interval. For each time interval we calculate JSD between daily patterns of egos $i$ and $j$: 
$d_\mathrm{ref}^{ij} = \frac{1}{3}\left(d_{11}^{ij}+d_{22}^{ij}+d_{33}^{ij}\right)$. As seen in Fig.~\ref{fig:JSD},  $d_\mathrm{self}$ takes on average lower values than $d_\mathrm{ref}$, meaning that there is more similarity between an ego's consecutive daily patterns than between the patterns of different egos in one interval. On average, for each ego, $87\%\pm 12\%$ of reference distances are higher than self-distances. Comparing average values of distances over all egos we get $ \langle d_\mathrm{ref} \rangle = 0.083 \pm 0.28$ while $\langle d_\mathrm{self}\rangle = 0.05 \pm 0.22 \,(\langle d_\mathrm{self}\rangle < \langle d_\mathrm{ref}\rangle$ with $t = 6.98$ and $p \ll 10^{-6}$, two-sample unequal variance t-test). To validate these results with another method, we have used the $\ell^2$ norm (see Methods), and the results qualitatively agree with JSD: $\langle d_\mathrm{self} \rangle= 0.11 \pm  0.02$ and $\langle d_\mathrm{ref} \rangle= 0.14 \pm 0.03$ with $t= 6.11$, $p < 10^{-5}$.

\subsection{Alter-specificity in call patterns}

We next turn to the question of where the individual daily patterns come from, and study the extent of a social component -that is, alter-specificity- in the call patterns. One can conceive of two extreme cases: 1) The patterns are entirely endogenous and the rate of call activity at each hour of the day is intrinsic to the ego. In this case, the called alters are picked at random (however, with a weight proportional to the time-averaged fraction of calls to each alter). 2) The patterns are alter-specific, that is, calls to certain alters are placed at certain hours, and the daily pattern is a superposition of the alter-specific patterns. 

\begin{figure*}[t]
\includegraphics[width=0.7\textwidth]{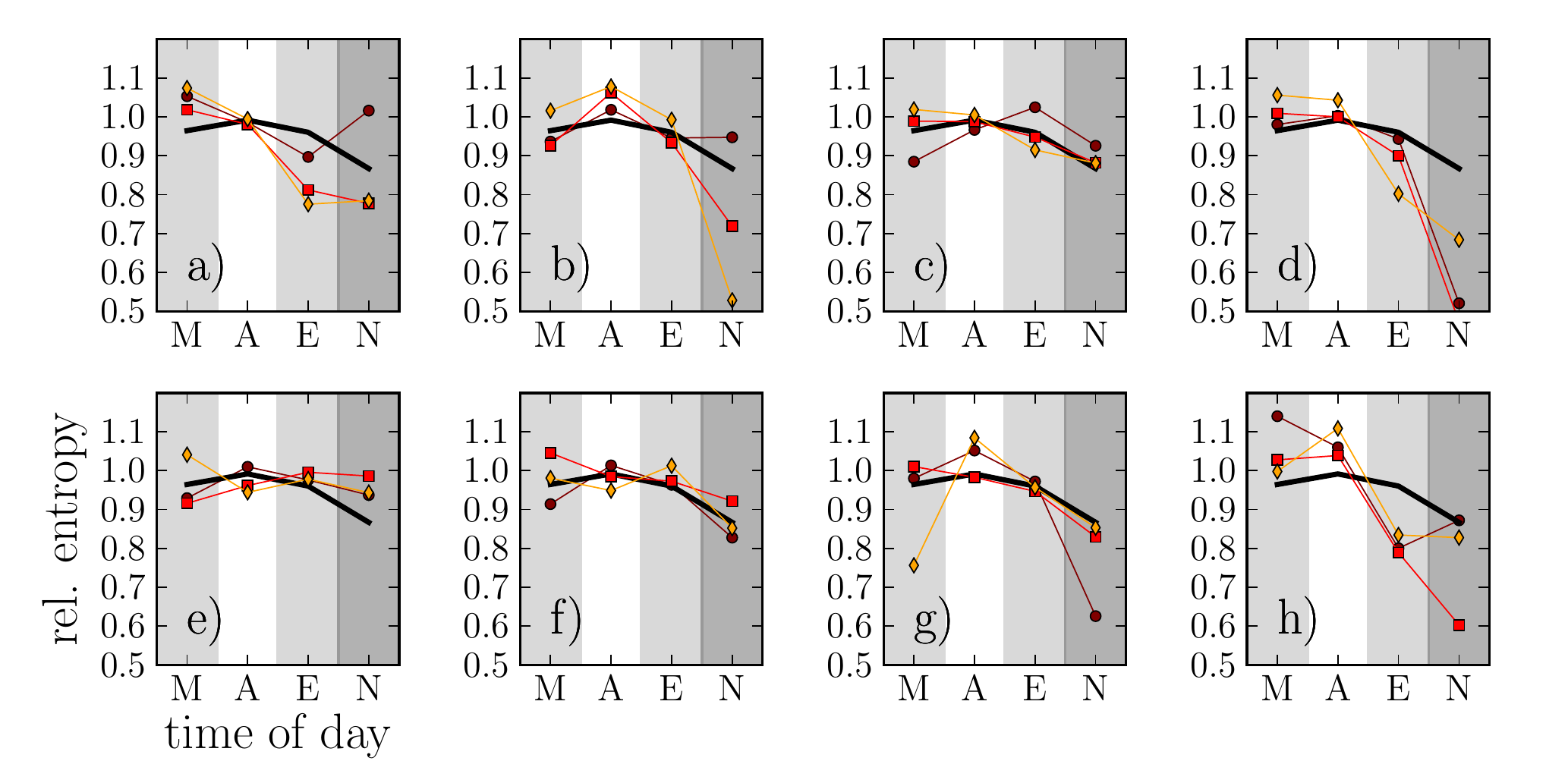}
\caption{The relative entropies for the same 8 individuals as in Fig1, calculated for 6-hour intervals (M: morning 6AM-12AM, A: afternoon 12AM-6PM, E: evening 6PM-0AM, N: night 0AM-6AM). ($\circ$): interval $I_1$, ($\square$): interval $I_2$, ($\diamond$): interval $I_3$. The black line indicates the average relative entropy for all 24 individuals over all three intervals.}
\label{fig:entropies}
\end{figure*}

To assess the extent of alter-specificity, we divide the day into 6-hour time spans (night: 0AM--6AM, morning: 6AM--12AM, afternoon: 12AM--6PM, evening: 6PM--12PM), and for each alter and each time span, compute the relative call entropies $H_{\mathrm{rel}}$. First, call entropies $H_{\mathrm{orig}}$ are calculated from the original data for each 6-hour span. To get the relative entropies $H_{\mathrm{rel}}$, these values are then divided by average entropies $\langle H_{\mathrm{ref}} \rangle$ calculated for a reference model where the times of calls to all alters are shuffled on a weekly basis for each ego (see Methods). If for a given 6-hour span $H_{\mathrm{rel}}<1$, calls to certain alters are emphasized within that time span, whereas if $H_{\mathrm{rel}}\approx 1$, there is no alter-specificity.

The relative entropies $H_{\mathrm{rel}}$ for the same 8 individuals as in Fig.~\ref{fig:daily_patterns} are displayed in Fig.~\ref{fig:entropies}, together with averaged relative entropy for all 24 individuals over all three intervals. The average relative entropy is at its highest in the afternoon, with $\langle H_{\mathrm{rel}}\rangle \approx 1$, indicating large diversity of called alters. $\langle H_{\mathrm{rel}}\rangle $ has its lowest point at night, when the number of calls is also low (see Fig.~\ref{fig:daily_patterns}). This indicates that the few calls made at night are typically directed to specific alters. As with the call frequency patterns, Fig.~\ref{fig:daily_patterns} clearly points out that the entropy patterns of different egos are different (compare, \emph{e.g.}, panels d and e). Likewise, each ego's patterns appear fairly persistent; however, there is more variation here, especially in the morning and at night when the call frequency is low and the entropy measures are as a result noisy. 

\begin{figure*}[!t]
\includegraphics[width=0.7\textwidth]{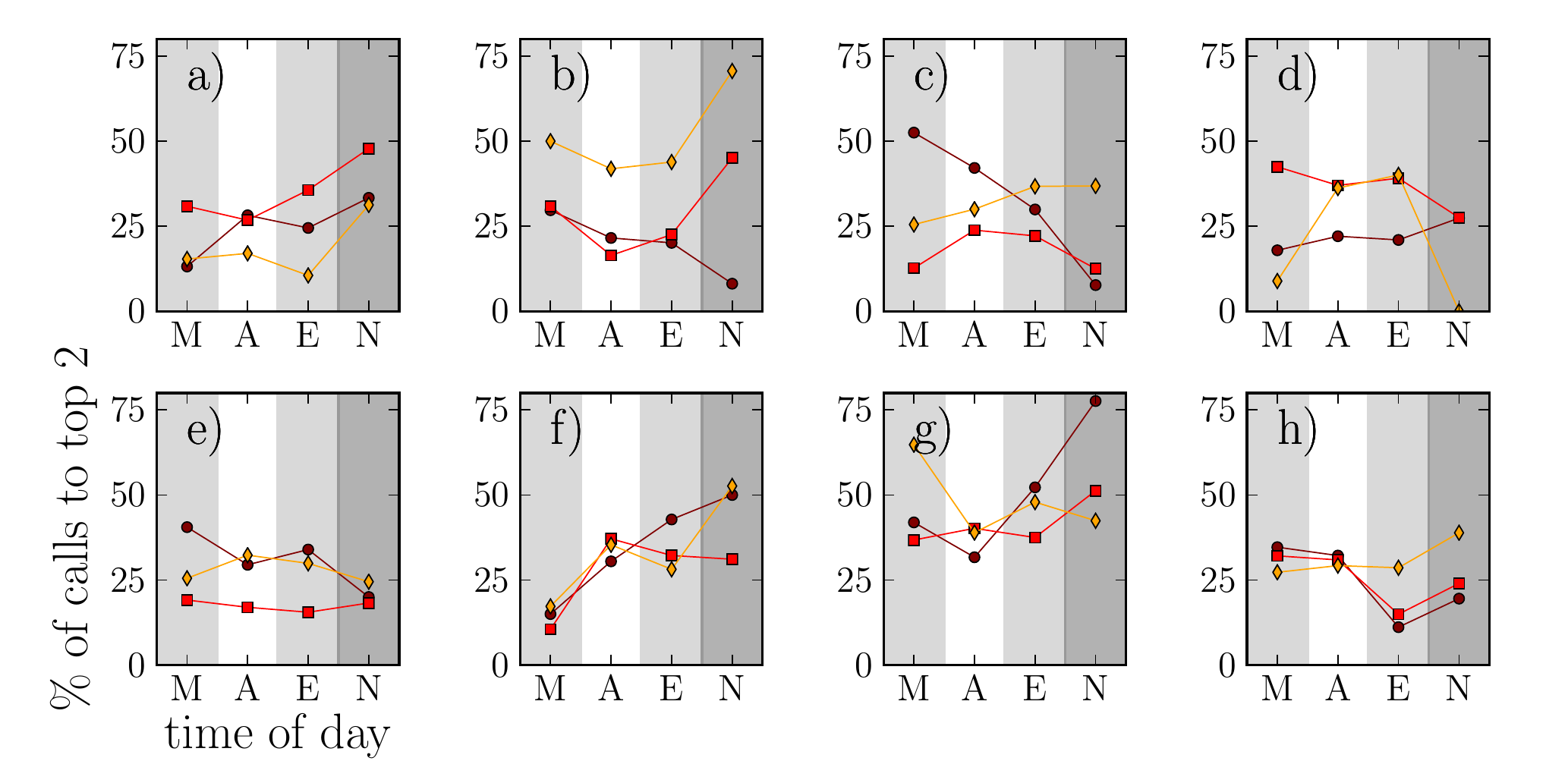}
\caption{The fractions of calls to the two top-ranked alters for the same 8 individuals as in Fig1, calculated for the same 6-hour intervals as in Fig3 (M, A, E, N). ($\circ$): interval $I_1$, ($\square$): interval $I_2$, ($\diamond$): interval $I_3$.}
\label{fig:top_fractions}
\end{figure*}

We next focus on the specific alters behind the low-entropy times of day. For this, we first count the total number of calls by each ego to each alter in each interval, and rank the alters of each ego according to this number.Alter ranks based on number of calls are known to reflect both the level of emotional closeness between ego and alter (as indexed on a standard psychological 1-10 emotional closeness scale), and the frequency face-to-face contact between ego and alter~\cite{Saramaki2014}. Then, for each 6-hour interval (morning, afternoon, evening, and night) we calculate the fraction of calls directed at the two top-ranked alters. These fractions are shown in Fig.~\ref{fig:top_fractions}, again for the same individuals as in Fig.~\ref{fig:daily_patterns}. On average, it appears that the fraction of calls to two top-ranked alters of each ego increases towards late hours and is often the highest at night, when there is in general only a small number of calls and low relative entropy. The high fractions indicate that decrease of entropy towards night typically comes from calls mainly to top-ranked alters; note that again, there is individual variation and although the top-alter fractions are often similar across intervals, in some cases, interval $I_1$ behaves differently. This interval corresponds to the participants finishing high school and the following summer holidays, so differences in call behavior can be expected.

Because Figs.~\ref{fig:entropies} and~\ref{fig:top_fractions} point towards a correlation between low entropy and calls to top-ranked alters, we next quantify this as follows: As the baseline levels and slopes of Fig.~\ref{fig:top_fractions} have a lot of variation, we take each ego and their relative entropies and fractions of calls to top-ranked alters at each 6-hour interval. Then, we compute the Pearson correlation coefficient between entropies and top-alter fractions for all egos. Out of the resulting 24 correlation coefficients, 15 were significant with $p<0.05$, with two positive coefficients and 13 negative averaging at $r\approx -0.72$. Thus for more than half of the egos, low entropy is clearly associated with a high fraction of calls to top alters, while for almost all the rest, no conclusive results can be drawn.

Since there are alter-specific communication patterns and the nature of communication depends on the time of day, we also look at call durations 
at different times. Here, we use data on ego and alter attributes from the conducted surveys.
Previous studies have looked at gender differences in talkativeness as well as differences in usage of phones(both for landlines and mobile phones)~\cite{Onnela2014, Stoica2010, Zainudeen2010}, using data from different countries and age groups. Most of the recent studies of talkativeness suggest that men and women are similar~\cite{Onnela2014, Mehl2007}. However in most studies which compare phone usage difference between men and women, women have been reported to have longer calls~\cite{Smoreda2000, Ling1998}. The differences in phone usage of males and females have been linked to their different social roles~\cite{Rakow1992, DeBaillon2005, Ling2004}. Here, we add two more dimensions and look at call durations at different times of the day, as well as durations of calls to different types of social links (kin or friend/acquaintance). 

\begin{figure}[!t]
\includegraphics[width=0.85\linewidth]{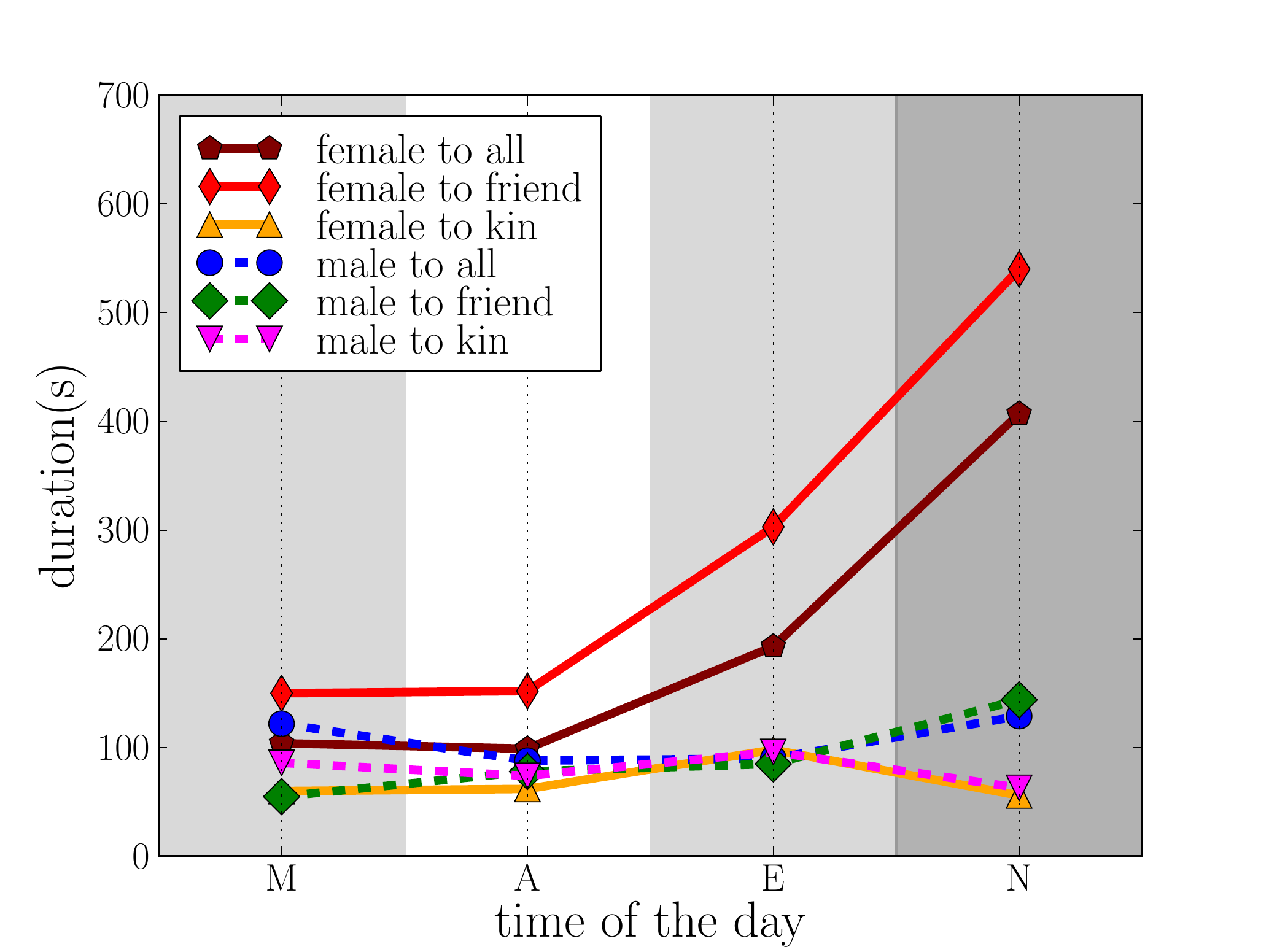}
\caption{Average duration of calls made by males and females to their kin, friends, and all social contacts.}
\label{fig:duration1}
\end{figure}

\begin{figure}[!t]
\includegraphics[width=0.85\linewidth]{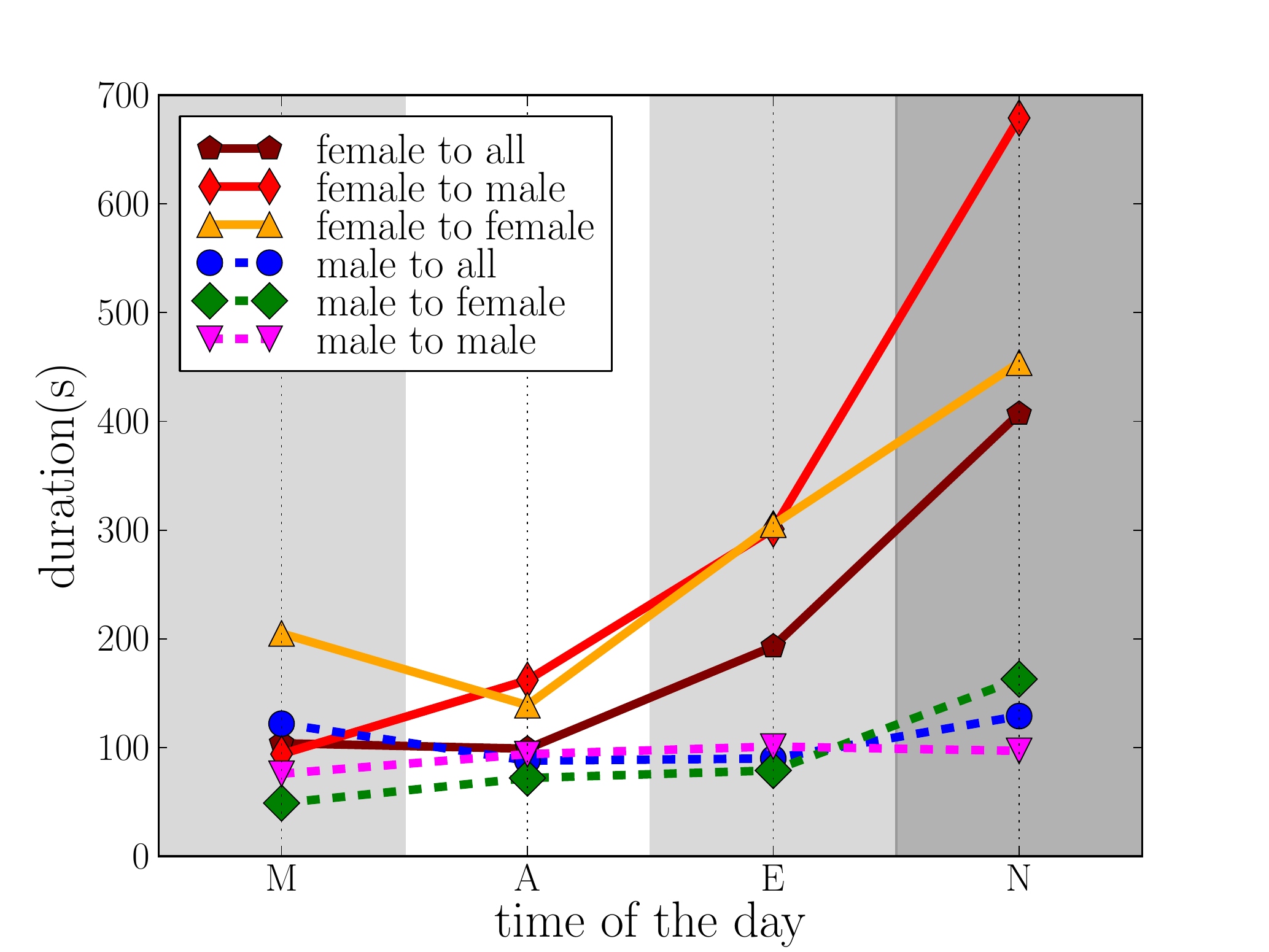}
\caption{Comparison of the average duration of calls to social contacts of the same and opposite gender, separately for females and males.} 
\label{fig:duration2}
\end{figure}

Fig.~\ref{fig:duration1} shows that overall, the average durations of calls by females are longer than those of calls by males, and that the difference largely depends on the time of day such that it increases towards the evening and is highest at night. A closer look shows that this difference arises mostly from calls to friends. Male and female call durations to kin are fairly similar and do not depend much on the time of day. When the gender of the called alters is analysed (Fig.~\ref{fig:duration2}), it is seen that by far the longest calls are by female egos to male alters at night; again the differences are the smallest in the afternoon, \emph{i.e.} when all egos are typically in a similar social setting (at school, work, or university). The finding  agrees with previous studies which suggest that females have different bonding strategies and use phones for different purposes compared to men~\citep{DeBaillon2005,Sarch1993}. Since nighttime calls are often targeted at top-ranking alters (who typically are emotionally close~\cite{Saramaki2014}), and the egos are in their late teens and are possibly experiencing emotionally intense relationships with their romantic partners, it is likely that these long calls often relate to romantic relationships.\\
Since we have a relatively small sample of individuals (24 total), one might think that the high values for call durations for females in afternoons and nights might only be caused by one or few females who make very long calls in those hours. To rule out this possibility, for each individual we compare call durations made in the morning or afternoon with duration of calls made in the afternoon or night, using two-sample unequal variance t-test. We see that for 9 out of 12 females p-values for this test are less than $10^{-6}$, whereas only for 2 males out of 12 we have such small p-values.   
%
%
%
%

\section{Discussion}

In contrast to conventional studies on daily patterns and circadian rhythms in social networks that focus on aggregates of very large numbers of individuals, here we focused on a small but rich sample that combined questionnaire and mobile phone data
in order to be able to explore in much greater detail features that characterise individualsÕ circadian rhythms. Our focus has been on three specific issues, namely (1) whether there are individual-specific patterns of calling that mirror previously demonstrated individual patterns in the way individuals allocate their social capital to their alters; (2) whether the individual-specific patterns of calling are persistent in light of network turnover and (3) whether there are gender differences in calling patterns.  We show that individuals do indeed have different 
daily patterns of call activity. These patterns vary beyond simple morningness/eveningness, as measured in questionnaire studies, and appear to be characteristic of the individual, in much the same way as their characteristic way of distributing their social capital among their alters~\cite{Saramaki2014}. Thus these individual patterns are persistent, in that the pattern of distributing calls across the day is consistent across the three time periods, despite the high degree of network turnover in the 18 months of the study, associated with leaving school and entering work or University~\cite{Roberts2011,Saramaki2014}. We also showed that there are striking gender differences in call duration pattern across the day: while women's calls are generally longer than men's calls, this was especially true during the evening and at night. Evening calls to males and to friends by female egos were especially long, and often involved calls to specific individuals, usually the top-ranked alters, who may be boyfriends.

	Given that humans naturally spend the night asleep, the tendency for calls to exhibit a striking diurnal periodicity is not, of itself, especially surprising, of course. However, in our sample, the vast majority of calls were made between midday and late evening, with the bulk of these occurring in the 6-hour slot between 12AM-6PM Fig.~\ref{fig:daily_patterns}. It is notable that, despite our essentially diurnal nature, rather few calls were made before midday. Most calls made by our cohort of subjects are likely to have been purely social rather than functional (i.e. work or leisure-activity related). Weissner ~\cite{Wiessner2014} reported that certain types of conversations (notably story-telling and social conversations) are much more common during the evening than during the day among !Kung San hunter gatherers, with conversations involving economic matters or social criticism taking place mainly during daylight hours. In her sample, 81\% of evening fireside conversations involved storytelling (relaying of adventures or experiences, especially in far off places, or tales about myths, social conventions and rituals, experiences during trance states or real life travels). Casual observation suggests that we find social events in the evening more engaging than ones held during the day, and that stories (and especially stories of the supernatural) told at night have an especial frisson. Our data suggest that this nocturnal intensity spills over into conversations between close friends (but, interestingly, much less often between close kin). Why this should be remains unclear, and would clearly merit more detailed study.
	
	Within this broad pattern, the individual differences in the distribution of calling, and particularly the persistence of these individual differences in the light of social network turnover, are strongly suggestive of some kind of personality characteristic. It is possible that these differences in personal style simply reflect individual differences in circadian pattern~\cite{Zhao2014, Tsaousis2010, Keren2010}, and are a consequence of the fact that some individuals are more active in the morning and others more active in the evening. It is perhaps less likely that the calling patterns are due to differences in individual sleep/wake cycles, since the demands of the working day are likely to have required everyone to be active in the morning and even early risers are unlikely to have gone to bed by 6PM (note that circumstances such as unemployment do have effects on daily cycles of individuals, see~\cite{Llorente2014}). However, it could be that, physiologically, morning people are more likely to feel motivated to be socially engaged in the daytime and evening people more likely to be so in the evening. The fact that some individuals find the evening hours particularly attractive, while others prefer the day, remains intriguing in this context and obviously merits more detailed investigation. 
	
	Notwithstanding the fact that some individuals are night-oriented and others day-oriented, it seems that many (though not all) egos prefer to call certain alters at night. These are typically the one or two individuals (mainly males and friends) that have special status for the ego (Fig.~\ref{fig:top_fractions}). We know from our detailed questionnaire data that the individuals who egos call most often are those to whom they are emotionally closest, and those they have the most frequent face-to-face contact with~\cite{Saramaki2014}. It seems that this is especially characteristic of female egos, and much less so of male egos. Unlike women, men do not call either their girlfriends or their same-sex best friends for long chats in the evenings (even though their girlfriends may call them). This striking sex difference in whom actively makes the effort to call is reminiscent of the finding reported by Palchykov et al.~\cite{palchykov2012}, for a very large cellphone dataset, that younger women (in particular) are much more proactive in calling their primary male contact than are men. As such, our finding speaks to the importance of female choice in human mating strategies -- that having made up their mind, women are typically much more focused in pursuing and investing in their relationships, and especially romantic relationships. This striking difference between the two sexes may reflect women's more intensely social nature compared to men. The fact that these "special" calls are reserved for the evening reinforces the suggestion that the hours of darkness have a special quality for certain kinds of social interactions and social relationships. 
	
For both sexes, it was much less common to call kin during the evening. This would reinforce the claim that relationships with kin are less fragile than those with friends, and hence require less persistent and less special servicing~\cite{Roberts2011}. Reserving calls to these individuals for times of the day when they are, or might seem to be, more intimate may reinforce the sense that the relationship is special. In effect, kin relationships come for free by virtue of the fact that they are kin and ego is embedded in a densely interconnected web of relationships with them, and therefore require less active maintenance. In contrast, the quality of friendships deteriorates rapidly (within months) in the absence of sufficiently frequent contact~\cite{Oswald2003, Cummings2006, Roberts2011}.

	Our study was, by the standards of most online network analyses, very small scale, of course. However, its merit is that we have a complete record of the identity of all Alters called by our subjects, as well as their relationships with those alters. We know that the individuals they call most often are those with whom they have the emotionally closest relationships~\cite{Saramaki2014}, and those with whom they are most socially active ~\cite{Roberts2011}. In this respect, we are able to add a layer of detail that is not usually possible in large scale network studies.
	
	A strength of our study was that it combined detailed mobile phone records with questionnaire data. Thus we have information on the nature of the relationship between egos and the alters they are calling, in terms of gender, kinship and emotional intensity. Further, we know from previous findings that the number and duration of phone calls relates to the emotional intensity of the relationship, as well as the level of social activity~\cite{Roberts2011,Saramaki2014}. This dataset therefore allows for analysis of the social nature of circadian rhythms, rather than simply examining aggregate analyses of mobile phone activity or broad scale questionnaire data. Due to the intensive, longitudinal data of the data collection, the sample size is relatively small. However, the nature of the data allows us to add a level of individual detail on identity of the callers and callees that is usually not possible in large scale social network studies

\section{Competing interests}
We have no competing interests.

\section{Author contributions}
SGBR and RD designed and performed the experiment and data collection. EL parsed and pre-processed data. TA, JS, SGBR, RD, EL, EM, and FTS conceived research and designed data analysis. TA and JS performed data analysis. TA, JS, EM, SGBR, and RD drafted the manuscript.

\section{Acknowledgements}
TA would like to thank Juan Perotti and Hang-Hyun Jo for useful discussions at different stages of the work. TA and JS acknowledge support from the Academy of Finland, project n:o 260427 and the computational resources provided by Aalto Science-IT project. RD's research is supported by an ERC Advanced grant. The collection of the data by SGBR and RD was made possible by a grant from the UK EPSRC and ESRC research councils.

\appendix	
\section*{Appendix}\label{Methods}

\subsection{Our data and its use}

Our dataset includes 18 months of (outgoing) call and text records of the 24 individuals. In this study we have only used call records (both to mobile phones and landlines). Participants filled out 3 questionnaires in the beginning (month 0), in the middle (month 9) and at the end (month 18), about people in their social network. They identified each contact (alter) as kin or non-kin, and provided all the different phone numbers that one contact might possibly have, and therefore these records are very comprehensive and do not miss a part of communication because an alter has several numbers (with multiple phone providers) and/ or uses landline. They also provided other information such as gender, how emotionally close they are to the person and the frequency of face-to-face contact with the person. The data on phone calls was obtained from the fully time-stamped, itemised monthly bills provided by arrangement with the provider, with the agreement of the subscriber. Participants received free mobile phone accounts for 18 months in return for taking part in the study. The participants, questionnaire and mobile phone datasets are described fully in previous publications~\cite{Roberts2011,Saramaki2014}. Anonymized, aggregated data is available online (see the SI of ~\cite{Saramaki2014}).

\subsection{Calculating daily patterns}
To calculate daily patterns of each ego, we have taken data from all days in the time interval of interest and have allocated each call to a 6-hour time bin based on its time stamp. We then count total number of events of each hour and divide it by total number of events (of that ego) during the the time interval, to get the fraction of calls in that hour. In each time interval, we only used data of complete weeks in that time interval. 
\subsection{Measuring similarity of patterns}

The Jensen-Shannon divergence is a measure of the difference of two probability distributions. It is a form of Kullback-Leibler divergence (KLD); unlike KLD, it works for probability distributions that contain zero-valued elements. The JSD for two probability discrete distributions $P_1$ and $P_2$ follows the formula 
$JSD(P_1,P_2) = H(\frac{1}{2}P_1+\frac{1}{2}P_2)-\frac{1}{2}[H(P_1)-H(P_2)]$, where $P_i={p_i(a)}$ and $p_i(a)$ is the fraction of calls to each alter, and $H$ is the Shannon entropy ($H(P)=-\sum p(a)\log p(a)$).

We have also used the $l^2$-norm as a way to verify our results calculated using JSD. $\ell^2$-norm is a similarity measure of two distributions, which is defined as: $\ell^2=\sqrt{\sum p_1(a)-p_2(a)|^2} $ . 
\subsection{Entropy patterns and relative entropies}

We calculate the call entropy for a given hour (or range of hours) as follows: first, the fraction of calls out of all calls  to each alter $a$, $p_a$, is counted for the specified hour (range of hours). Then, the call entropy for this hour (range of hours) is computed as $H_{\mathrm{orig}}=-\sum_a p_a\log p_a$. In order to obtain the relative entropy, we repeatedly shuffle the original data as follows: for each week,  the times and recipient alters of all calls are randomly shuffled. This reference model corresponds to a situation where the original call frequency pattern and the number of calls to each alter are the same, but no preference is shown to any specific alter at any specific time. Then, for each shuffled set of data, we calculate call entropy similarly as for the original data, and average over $N=1,000$ realizations to get the average reference entropy $\langle H_{\mathrm{ref}}\rangle$. Finally, the relative entropy is obtained as $H_{\mathrm{rel}}=H_{\mathrm{orig}}/\langle H_{\mathrm{ref}} \rangle$. The shuffling for the reference model is done on a weekly basis in order to minimize the effects of long-term dynamics, such as declining numbers of calls to alters, or alters appearing for the first time within the studied 6-month interval.

\bibliography{circadian} 

\def\url#1{}
\begin{thebibliography}{10}
\expandafter\ifx\csname url\endcsname\relax
  \def\url#1{\texttt{#1}}\fi
\expandafter\ifx\csname urlprefix\endcsname\relax\def\urlprefix{URL }\fi
\providecommand{\bibinfo}[2]{#2}
\providecommand{\eprint}[2][]{\url{#2}}

\bibitem{Kerkhof1985}
\bibinfo{author}{Kerkhof, G.~A.}
\newblock \bibinfo{title}{Inter-individual differences in the human circadian
  system: a review}.
\newblock \emph{\bibinfo{journal}{Biological Psychology}}
  \textbf{\bibinfo{volume}{20}}, \bibinfo{pages}{83--112}
  (\bibinfo{year}{1985}).

\bibitem{Czeisler1999}
\bibinfo{author}{Czeisler, C.~A.} \emph{et~al.}
\newblock \bibinfo{title}{Stability, precision, and near-24-hour period of the
  human circadian pacemaker}.
\newblock \emph{\bibinfo{journal}{Science}} \textbf{\bibinfo{volume}{284}},
  \bibinfo{pages}{2177--2181} (\bibinfo{year}{1999}).

\bibitem{Panda2002}
\bibinfo{author}{Panda, S.}, \bibinfo{author}{Hogenesch, J.~B.} \&
  \bibinfo{author}{Kay, S.~A.}
\newblock \bibinfo{title}{Circadian rhythms from flies to humans}.
\newblock \emph{\bibinfo{journal}{Nature}} \textbf{\bibinfo{volume}{417}},
  \bibinfo{pages}{329--335} (\bibinfo{year}{2002}).

\bibitem{Zhao2014}
\bibinfo{author}{Zhao, R.} \emph{et~al.}
\newblock \bibinfo{title}{Influences of age, gender, and circadian rhythm on
  deceleration capacity in subjects without evident heart diseases}.
\newblock \emph{\bibinfo{journal}{Annals of Noninvasive Electrocardiology}}
  (\bibinfo{year}{2014}).

\bibitem{Tsaousis2010}
\bibinfo{author}{Tsaousis, I.}
\newblock \bibinfo{title}{Circadian preferences and personality traits: A
  meta-analysis}.
\newblock \emph{\bibinfo{journal}{European Journal of Personality}}
  \textbf{\bibinfo{volume}{24}}, \bibinfo{pages}{356--373}
  (\bibinfo{year}{2010}).
\newblock \urlprefix\url{http://dx.doi.org/10.1002/per.754}.

\bibitem{Keren2010}
\bibinfo{author}{Keren, H.}, \bibinfo{author}{Boyer, P.},
  \bibinfo{author}{Mort, J.} \& \bibinfo{author}{Eilam, D.}
\newblock \bibinfo{title}{Pragmatic and idiosyncratic acts in human everyday
  routines: The counterpart of compulsive rituals}.
\newblock \emph{\bibinfo{journal}{Behavioural Brain Research}}
  \textbf{\bibinfo{volume}{212}}, \bibinfo{pages}{90 -- 95}
  (\bibinfo{year}{2010}).

\bibitem{Preti2001}
\bibinfo{author}{Preti, A.} \& \bibinfo{author}{Miotto, P.}
\newblock \bibinfo{title}{Diurnal variations in suicide by age and gender in
  italy}.
\newblock \emph{\bibinfo{journal}{Journal of Affective Disorders}}
  \textbf{\bibinfo{volume}{65}}, \bibinfo{pages}{253 -- 261}
  (\bibinfo{year}{2001}).

\bibitem{Kouchaki2014}
\bibinfo{author}{Kouchaki, M.} \& \bibinfo{author}{Smith, I.~H.}
\newblock \bibinfo{title}{The morning morality effect: The influence of time of
  day on unethical behavior}.
\newblock \emph{\bibinfo{journal}{Psychological Science}}
  \textbf{\bibinfo{volume}{25}}, \bibinfo{pages}{95--102}
  (\bibinfo{year}{2014}).

\bibitem{Refinetti2005}
\bibinfo{author}{Refinetti, R.}
\newblock \bibinfo{title}{Time for sex: nycthemeral distribution of human
  sexual behavior}.
\newblock \emph{\bibinfo{journal}{Journal of Circadian Rhythms}}
  \textbf{\bibinfo{volume}{3}} (\bibinfo{year}{2005}).

\bibitem{Hu2004}
\bibinfo{author}{Hu, K.} \emph{et~al.}
\newblock \bibinfo{title}{Endogenous circadian rhythm in an index of cardiac
  vulnerability independent of changes in behavior}.
\newblock \emph{\bibinfo{journal}{Proceedings of the National Academy of
  Sciences}} \textbf{\bibinfo{volume}{101}}, \bibinfo{pages}{18223--18227}
  (\bibinfo{year}{2004}).

\bibitem{Song2010}
\bibinfo{author}{Song, C.}, \bibinfo{author}{Qu, Z.}, \bibinfo{author}{Blumm,
  N.} \& \bibinfo{author}{Barab{\'a}si, A.-L.}
\newblock \bibinfo{title}{Limits of predictability in human mobility}.
\newblock \emph{\bibinfo{journal}{Science}} \textbf{\bibinfo{volume}{327}},
  \bibinfo{pages}{1018--1021} (\bibinfo{year}{2010}).

\bibitem{Yasseri2012}
\bibinfo{author}{Yasseri, T.}, \bibinfo{author}{Sumi, R.} \&
  \bibinfo{author}{Kert\'{e}sz, J.}
\newblock \bibinfo{title}{Circadian patterns of wikipedia editorial activity: A
  demographic analysis}.
\newblock \emph{\bibinfo{journal}{PLoS One}} \textbf{\bibinfo{volume}{7}},
  \bibinfo{pages}{e30091} (\bibinfo{year}{2012}).

\bibitem{Yasseri2013}
\bibinfo{author}{Yasseri, T.}, \bibinfo{author}{Quattrone, G.} \&
  \bibinfo{author}{Mashhadi, A.}
\newblock \bibinfo{title}{Temporal analysis of activity patterns of editors in
  collaborative mapping project of openstreetmap}.
\newblock In \emph{\bibinfo{booktitle}{Proceedings of the 9th International
  Symposium on Wikis and Open Collaboration}} (\bibinfo{year}{2013}).

\bibitem{Thij2014}
\bibinfo{author}{ten Thij, M.}, \bibinfo{author}{Bhulai, S.} \&
  \bibinfo{author}{Kampstra, P.}
\newblock \bibinfo{title}{Circadian patterns in twitter}.
\newblock In \emph{\bibinfo{booktitle}{DATA ANALYTICS 2014, The Third
  International Conference on Data Analytics}}, \bibinfo{pages}{12--17}
  (\bibinfo{year}{2014}).

\bibitem{Ho2012}
\bibinfo{author}{Ho, H.-H.}, \bibinfo{author}{Karsai, M.},
  \bibinfo{author}{Kert\'{e}sz, J.} \& \bibinfo{author}{Kaski, K.}
\newblock \bibinfo{title}{Circadian pattern and burstiness in mobile phone
  communication}.
\newblock \emph{\bibinfo{journal}{New Journal of Physics}}
  \textbf{\bibinfo{volume}{14}}, \bibinfo{pages}{013055}
  (\bibinfo{year}{2012}).

\bibitem{Louail2014}
\bibinfo{author}{Louail, T.} \emph{et~al.}
\newblock \bibinfo{title}{From mobile phone data to the spatial structure of
  cities}.
\newblock \emph{\bibinfo{journal}{arXiv:1401.4540 [physics.soc-ph]}}
  (\bibinfo{year}{2014}).

\bibitem{Roberts2011}
\bibinfo{author}{Roberts, S. G.~B.} \& \bibinfo{author}{Dunbar, R. I.~M.}
\newblock \bibinfo{title}{The costs of family and friends: an 18-month
  longitudinal study of relationship maintenance and decay}.
\newblock \emph{\bibinfo{journal}{Evolution and Human Behavior}}
  \textbf{\bibinfo{volume}{32}}, \bibinfo{pages}{186--197}
  (\bibinfo{year}{2011}).

\bibitem{Saramaki2014}
\bibinfo{author}{Saram\"{a}ki, J.} \emph{et~al.}
\newblock \bibinfo{title}{Persistence of social signatures in human
  communication}.
\newblock \emph{\bibinfo{journal}{Proc. Natl. Acad. Sci. U.S.A.}}
  \textbf{\bibinfo{volume}{111}}, \bibinfo{pages}{942--947}
  (\bibinfo{year}{2014}).

\bibitem{Onnela2014}
\bibinfo{author}{Onnela, J.-P.}, \bibinfo{author}{Waber, B.~N.},
  \bibinfo{author}{Pentland, A.}, \bibinfo{author}{Schnorf, S.} \&
  \bibinfo{author}{Lazer, D.}
\newblock \bibinfo{title}{Using sociometers to quantify social interaction
  patterns}.
\newblock \emph{\bibinfo{journal}{Sci. Rep.}} \textbf{\bibinfo{volume}{4}}
  (\bibinfo{year}{2014}).

\bibitem{Stoica2010}
\bibinfo{author}{Stoica, A.}, \bibinfo{author}{Smoreda, Z.},
  \bibinfo{author}{Prieur, C.} \& \bibinfo{author}{Guillaume, J.-L.}
\newblock \bibinfo{title}{Age, gender and communication networks}.
\newblock \emph{\bibinfo{journal}{NetMob - Ananlysis of Mobile Phone Networks
  2010 - Communication Proposal}}  (\bibinfo{year}{2010}).

\bibitem{Zainudeen2010}
\bibinfo{author}{Zainudeen, A.}, \bibinfo{author}{Iqbal, T.} \&
  \bibinfo{author}{Samarajiva, R.}
\newblock \bibinfo{title}{Who's got the phone? gender and the use of the
  telephone at the bottom of the pyramid}.
\newblock \emph{\bibinfo{journal}{New Media and Society}}
  \textbf{\bibinfo{volume}{12}}, \bibinfo{pages}{549--566}
  (\bibinfo{year}{2010}).

\bibitem{Mehl2007}
\bibinfo{author}{Mehl, M.~R.}, \bibinfo{author}{Vazire, S.},
  \bibinfo{author}{Ramírez-Esparza, N.}, \bibinfo{author}{Slatcher, R.~B.} \&
  \bibinfo{author}{Pennebaker, J.~W.}
\newblock \bibinfo{title}{Are women really more talkative than men?}
\newblock \emph{\bibinfo{journal}{Science}} \textbf{\bibinfo{volume}{317}},
  \bibinfo{pages}{82} (\bibinfo{year}{2007}).

\bibitem{Smoreda2000}
\bibinfo{author}{Smoreda, Z.} \& \bibinfo{author}{Licoppe, C.}
\newblock \bibinfo{title}{Gender-specific use of the domestic telephone}.
\newblock \emph{\bibinfo{journal}{Social Psychology Quarterly}}
  \textbf{\bibinfo{volume}{63}}, \bibinfo{pages}{238--252}
  (\bibinfo{year}{2010}).

\bibitem{Ling1998}
\bibinfo{author}{Ling, R.}
\newblock \bibinfo{title}{"she calls, but it is for both of us you know": The
  use of traditional fixed and mobile telephony for social networking among
  norwegian parents}.
\newblock \bibinfo{type}{Tech. Rep.}, \bibinfo{institution}{Telenor R\&D}
  (\bibinfo{year}{1998}).

\bibitem{Rakow1992}
\bibinfo{author}{Rakow, L.}
\newblock \emph{\bibinfo{title}{Gender on the line : women, the telephone, and
  community life / Lana F. Rakow}} (\bibinfo{publisher}{University of Illinois
  Press Urbana}, \bibinfo{year}{1992}).

\bibitem{DeBaillon2005}
\bibinfo{author}{DeBaillon, L.} \& \bibinfo{author}{Rockwell, P.}
\newblock \bibinfo{title}{Gender and student-status differences in cellular
  telephone use}.
\newblock \emph{\bibinfo{journal}{International Journal of Mobile
  Communications}} \textbf{\bibinfo{volume}{3}}, \bibinfo{pages}{82--98}
  (\bibinfo{year}{2005}).

\bibitem{Ling2004}
\bibinfo{author}{Ling, R.}
\newblock \emph{\bibinfo{title}{The Mobile Connection : The Cell Phone's Impact
  on Society}} (\bibinfo{publisher}{Morgan Kaufmann}, \bibinfo{year}{2004}).

\bibitem{Sarch1993}
\bibinfo{author}{Sarch, A.}
\newblock \bibinfo{title}{Making the connection: Single women's use of the
  telephone in dating relationships with men}.
\newblock \emph{\bibinfo{journal}{Journal of Communication}}
  \textbf{\bibinfo{volume}{43}}, \bibinfo{pages}{128--144}
  (\bibinfo{year}{1993}).

\bibitem{Wiessner2014}
\bibinfo{author}{Wiessner, P.~W.}
\newblock \bibinfo{title}{Embers of society: Firelight talk among the
  ju/’hoansi bushmen}.
\newblock \emph{\bibinfo{journal}{Proc. Natl. Acad. Sci. U.S.A.}}
  \textbf{\bibinfo{volume}{111}}, \bibinfo{pages}{14027--14035}
  (\bibinfo{year}{2014}).

\bibitem{Llorente2014}
\bibinfo{author}{Llorente, A.}, \bibinfo{author}{Garcia-Herranz, M.},
  \bibinfo{author}{Cebrian, M.} \& \bibinfo{author}{Moro, E.}
\newblock \bibinfo{title}{Social media fingerprints of unemployment}.
\newblock \emph{\bibinfo{journal}{arXiv:1411.3140 [physics.soc-ph]}}
  (\bibinfo{year}{2014}).

\bibitem{palchykov2012}
\bibinfo{author}{Palchykov, V.}, \bibinfo{author}{Kaski, K.},
  \bibinfo{author}{Kert\'esz, J.}, \bibinfo{author}{Barab\'asi, A.-L.} \&
  \bibinfo{author}{Dunbar, R.}
\newblock \bibinfo{title}{Sex differences in intimate relationships}.
\newblock \emph{\bibinfo{journal}{Sci. Rep.}} \textbf{\bibinfo{volume}{2}}
  (\bibinfo{year}{2012}).

\bibitem{Oswald2003}
\bibinfo{author}{Oswald, D.~L.} \& \bibinfo{author}{Clark, E.~M.}
\newblock \bibinfo{title}{Best friends forever?: High school best friendships
  and the transition to college}.
\newblock \emph{\bibinfo{journal}{Personal Relationships}}
  \textbf{\bibinfo{volume}{10}}, \bibinfo{pages}{187--196}
  (\bibinfo{year}{2003}).

\bibitem{Cummings2006}
\bibinfo{author}{Cummings, J.}, \bibinfo{author}{Lee, J.} \&
  \bibinfo{author}{Kraut, R.}
\newblock \emph{\bibinfo{title}{{Communication technology and friendship during
  the transition from high school to college}}} (\bibinfo{publisher}{Oxford
  University Press.}, \bibinfo{year}{2006}).

\end{thebibliography}

\end{document}